\documentclass[a4paper]{PoS}
\pdfoutput=1
\usepackage{amsmath}

\title{Leading and next to leading large $n_f$ terms\\in the cusp anomalous dimension\\and the quark--antiquark potential}
\ShortTitle{Leading and next to leading large $n_f$ terms}
\author{\speaker{Andrey Grozin}%
\thanks{Preprint number MITP/16-044}\\
Mainz Institute of Theoretical Physics, Mainz University, Germany\\
Budker Institute of Nuclear Physics, Novosibirsk, Russia\\
E-mail: \email{A.G.Grozin@inp.nsk.su}}
\abstract{I discuss 3 related quantities: the cusp anomalous dimension,
the HQET heavy-quark field anomalous dimension, and the quark--antiquark potential.
Leading large $n_f$ terms can be calculated to all orders in $\alpha_s$.
Next to leading terms with the abelian color structure $C_F^2$ also can be found to all orders
(but not non-abelian $C_F C_A$ terms).
This talk is based on Appendices C and D in~\cite{Grozin:2015kna}.}
\FullConference{Loops and Legs in Quantum Field Theory\\
24--29 April 2016\\
Leipzig, Germany}

\begin{document}

\section{Introduction}
\label{S:Intro}

The one-loop cusp anomalous dimension
\begin{equation}
\Gamma(\alpha_s,\varphi) = C_F \frac{\alpha_s}{\pi} \left( \varphi \coth\varphi - 1 \right)
\label{Intro:Gamma1}
\end{equation}
follows from the soft radiation function in classical electrodynamics:
when a charge suddenly changes its velocity, it emits electromagnetic waves;
integrating the intensity over directions, one obtains~\cite{LL}
$\varphi \coth\varphi - 1$.
This result is probably known for more than 100 years,
and should be included in The Guinness Book of Records as the anomalous dimension known for a longest time.
The two-loop term has been calculated 30 years ago~\cite{Korchemsky:1987wg}
(and rewritten via $\mathop{\mathrm{Li}}\nolimits_{2,3}$ in~\cite{Kidonakis:2009ev}).
The three-loop term has been calculated recently~\cite{Grozin:2014axa,Grozin:2014hna,Grozin:2015kna}.

The HQET heavy-quark field anomalous dimension
(or the anomalous dimension of a straight Wilson line)
is known up to 3 loops.
At 2 loops, after a wrong calculation~\cite{Aoyama:1981ev},
the correct result has been obtained in~\cite{Knauss:1984rx},
and later in~\cite{Ji:1991pr,Broadhurst:1991fz,Broadhurst:1991fy,Gimenez:1991bf}.
The three-loop result has been obtained in~\cite{Melnikov:2000zc,Chetyrkin:2003vi}
(in the first paper~\cite{Melnikov:2000zc} it has been found as a by-product of the calculation
of the QCD on-shell heavy-quark field renormalization constant,
from the requirement that the QCD/HQET matching coefficient
for the heavy-quark field~\cite{Grozin:2010wa} is finite;
at 2 loops this has been done in~\cite{Broadhurst:1991fy}).

The quark--antiquark potential is known at two~\cite{Peter:1997,Schroder:1998vy}
and three~\cite{Smirnov:2008pn,Smirnov:2009fh,Anzai:2009tm} loops.

Some terms in perturbative series for these quantities can be obtained to all orders in $\alpha_s$.

\section{Large $n_f$ terms}

The terms with the highest power of $n_f$
at each order of perturbation theory for the cusp anomalous dimension $\Gamma$
have the structures $C_F (T_F n_f)^{L-1} \alpha_s^L$ ($L\ge1$).
They are known to all orders in $\alpha_s$.
The terms with next to highest power of $n_f$ have the structures
$C_F^2 (T_F n_f)^{L-2} \alpha_s^L$ and $C_F C_A (T_F n_f)^{L-2} \alpha_s^L$ ($L\ge3$).
The abelian ones (without $C_A$) can be also found to all orders in $\alpha_s$.
For this purpose it is sufficient to consider QED with $n_f$ massless lepton flavors:
$C_F=T_F=1$, $C_A=0$, $\beta_0=-\frac{4}{3}n_f$.
Let's introduce
\begin{equation}
b = \beta_0 \frac{\alpha}{4\pi}\,.
\label{Intro:b}
\end{equation}
We assume $b\sim1$ and take into account all powers of $b$;
$1/\beta_0\ll1$ is our small parameter,
and we consider only a few terms in expansions in $1/\beta_0$.

\begin{figure}[h]
\centering
\includegraphics{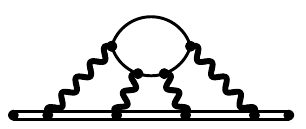}
\caption{The light-by-light diagram is $n_f \alpha^4$, and hence NNL$\beta_0$.}
\label{F:lbl}
\end{figure}

At the leading and next-to-leading large-$\beta_0$ orders (L$\beta_0$ and NL$\beta_0$),
the coordinate-space Wilson line of any shape is equal to
\begin{equation}
\log W = \raisebox{-6.25mm}{\begin{picture}(23,14.5)
\put(11.5,9.875){\makebox(0,0){\includegraphics{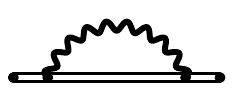}}}
\end{picture}}\,,
\label{Intro:exp}
\end{equation}
where the thick photon line is the full photon propagator with the NL$\beta_0$ accuracy.
This simple exponentiation formula is first broken at NNL$\beta_0$ order
by the light-by-light diagram (figure~\ref{F:lbl}).

With the NL$\beta_0$ accuracy the renormalization constant $Z$
of the heavy-to-heavy current (the cusp) is given by
\begin{equation}
\log W(t,t';\varphi) - \log W(t,t';0)
= \raisebox{-5.75mm}{\begin{picture}(23,12.5)
\put(11.5,6.25){\makebox(0,0){\includegraphics{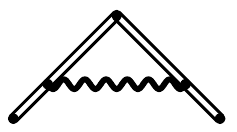}}}
\end{picture}}
- \raisebox{-11.5mm}{\begin{picture}(23,14.5)
\put(11.5,9.875){\makebox(0,0){\includegraphics{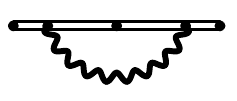}}}
\end{picture}}
= \log Z + \text{finite}
\label{Intro:ZW}
\end{equation}
(diagrams where both photon-interaction vertices are before the cusp,
or after the cusp, cancel in this difference).
Going to momentum space, we can express it via the vertex function $V(\omega,\omega';\varphi)$
(it is convenient to set $\omega'=\omega$,
in order to have a single-scale problem):
\begin{equation}
V(\omega,\omega;\varphi) - V(\omega,\omega;0)
= \raisebox{-4mm}{\begin{picture}(16,9)
\put(8,4.5){\makebox(0,0){\includegraphics{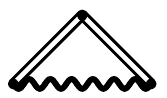}}}
\end{picture}}
- \raisebox{-11.5mm}{\begin{picture}(16,14.5)
\put(8,9.875){\makebox(0,0){\includegraphics{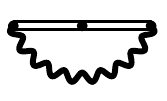}}}
\end{picture}}
= \log Z + \text{finite}\,.
\label{Intro:ZV}
\end{equation}
The HQET field renormalization can be obtained from $V(\omega,\omega;0)$.

\begin{figure}[h]
\centering
\includegraphics{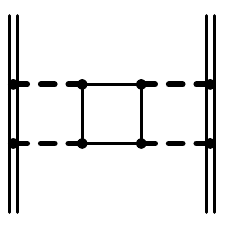}
\caption{The light-by-light diagram is $n_f \alpha^4$, and hence NNL$\beta_0$.}
\label{F:Vlbl}
\end{figure}

The static quark--antiquark potential can be considered similarly.
The terms with the highest power of $n_f$ in each order of perturbation theory
have the structures $C_F (T_F n_f)^L \alpha_s^{L+1}$ ($L\ge0$).
The terms with next to highest power of $n_f$ have the structures
$C_F^2 (T_F n_f)^{L-1} \alpha_s^{L+1}$ and $C_F C_A (T_F n_f)^{L-1} \alpha_s^{L+1}$ ($L\ge2$);
we'll consider only the abelian ones.
In the Coulomb gauge, up to NL$\beta_0$ the potential is given by the full Coulomb photon propagator
\begin{equation}
V(\vec{q}^{\,})
= \raisebox{-5.5mm}{\includegraphics{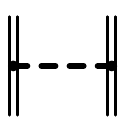}}
= - \frac{e_0^2}{\vec{q}^{\,2}} \frac{1}{1 - \Pi(-\vec{q}^{\,2})}
\label{Intro:V}
\end{equation}
($\Pi(q^2)$ is gauge invariant in QED,
and can be taken from covariant-gauge calculations).
This simple equality is first broken at NNL$\beta_0$ order
by the light-by-light diagram (figure~\ref{F:Vlbl}).

As discussed in~\cite{Grozin:2015kna}, conformal symmetry leads to the relation
between $\Gamma(\pi-\delta)$ at $\delta\to0$ and $V(\vec{q}^{\,})$:
\begin{equation}
\Delta \equiv \bigl[\delta \Gamma(\pi-\delta;\alpha_s)\bigr]_{\delta\to0}
- \frac{\vec{q}^{\,2} V(\vec{q};\alpha_s)}{4\pi} = 0
\label{Intro:Delta}
\end{equation}
(this relation has been observed in~\cite{Kilian:1993nk} at 2 loops).
In QCD (and QED) conformal symmetry is anomalous
(thus leading to non-zero $\beta$ function), and~\cite{Grozin:2015kna}
\begin{equation}
\Delta = \frac{\pi}{108} \beta_0 C_F \left(\frac{\alpha_s}{\pi}\right)^3 \left( 47 C_A - 28 T_F n_f \right)
+ \mathcal{O}(\alpha_s^4)\,.
\label{Intro:Delta3}
\end{equation}

\section{Leading $\beta_0$ order}
\label{S:Lb0}

The photon self energy at the L$\beta_0$ order is $\sim1$:
\begin{align}
&\Pi_0(k^2) = \raisebox{-5.5mm}{\includegraphics{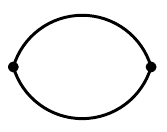}}
= \beta_0 \frac{e_0^2}{(4\pi)^{d/2}} e^{-\gamma\varepsilon} \frac{D(\varepsilon)}{\varepsilon} (-k^2)^{-\varepsilon}\,,
\nonumber\\
&D(\varepsilon) = e^{\gamma\varepsilon}
\frac{(1-\varepsilon) \Gamma(1+\varepsilon) \Gamma^2(1-\varepsilon)}{(1-2\varepsilon) (1-\frac{2}{3}\varepsilon) \Gamma(1-2\varepsilon)}
= 1 + \frac{5}{3} \varepsilon + \cdots
\label{Lb0:Pi}
\end{align}
The charge renormalization in the $\overline{\text{MS}}$ scheme is
\begin{equation}
\beta_0 \frac{e_0^2}{(4\pi)^{d/2}} e^{-\gamma\varepsilon} = b Z_\alpha(b) \mu^{2\varepsilon}\,.
\label{Lb0:MS}
\end{equation}
At the L$\beta_0$ order we can solve the RG equation
\begin{equation*}
\frac{d\log Z_\alpha}{d\log b} = - \frac{b}{\varepsilon+b}
\end{equation*}
and obtain
\begin{equation}
Z_\alpha = \frac{1}{1+b/\varepsilon}\,.
\label{Lb0:Za}
\end{equation}

\begin{figure}[b]
\begin{center}
\includegraphics{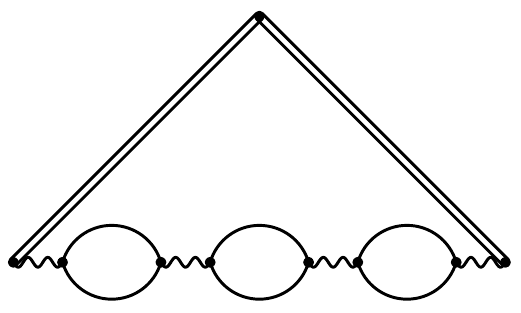}
\end{center}
\caption{The $L$-loop vertex diagram at the L$\beta_0$ order contains $L-1$ $\Pi_0$ insertions.}
\label{F:Lb0}
\end{figure}

The vertex $V(\omega,\omega;\varphi)$ is given by the one-loop diagram
with the factor $1/(1-\Pi(k^2))$ inserted in the integrand.
At the L$\beta_0$ order (figure~\ref{F:Lb0}) the result can be written in the form
\begin{equation}
V(\omega,\omega;\varphi)
= \raisebox{-4mm}{\begin{picture}(16,9)
\put(8,4.5){\makebox(0,0){\includegraphics{eb4.pdf}}}
\end{picture}}
= \frac{1}{\beta_0} \sum_{L=1}^\infty \frac{f(\varepsilon,L\varepsilon;\varphi)}{L} \Pi_0^L
+ \mathcal{O}\left(\frac{1}{\beta_0^2}\right)\,,
\label{Lb0:Vf}
\end{equation}
where $L$ is the number of loops and $\Pi_0$~(\ref{Lb0:Pi}) is taken at $-k^2=(-2\omega)^2$.
Reduction of such integrals to master ones,
as well as evaluation of these master integrals,
has been considered in~\cite{Grozin:2011rs}.
In Landau gauge we obtain
\begin{align}
&f(\varepsilon,u;\varphi) =
- \frac{(1-\frac{2}{3}\varepsilon) \Gamma(2-2\varepsilon) \Gamma(1-u) \Gamma(1+2u)}%
{(1-\varepsilon) \Gamma^2(1-\varepsilon) \Gamma(1+\varepsilon) \Gamma(2+u-\varepsilon)}
\nonumber\\
&{}\times\left[ \bigl((2+u-2\varepsilon)\cos\varphi-u\bigr)
\,_2F_1\left(\left.\begin{array}{c}1,1-u\\3/2\end{array}\right|\frac{1-\cos\varphi}{2}\right)
+ 1 \right]
\label{Lb0:fphi}
\end{align}
(in an arbitrary covariant gauge, a one-loop gauge-dependent contribution should be added).
The function $f(\varepsilon,u;\varphi)$ is regular at the origin:
\begin{equation}
f(\varepsilon,u;\varphi) = \sum_{n,m=0}^\infty f_{nm}(\varphi) \varepsilon^n u^m\,.
\label{Lb0:feuexp}
\end{equation}

The renormalization constant $Z$ can be written as
\begin{equation*}
\log Z = \frac{Z_1}{\varepsilon} + \frac{Z_2}{\varepsilon^2} + \cdots\,,\quad
Z_n = \mathcal{O}(b^n)\,.
\end{equation*}
Only $Z_1$ is needed in order to obtain
\begin{equation*}
\Gamma(b;\varphi) = - 2 \frac{d Z_1(b;\varphi)}{d\log b}\,;
\end{equation*}
higher $Z_n$ contain no new information,
and are uniquely reconstructed from $Z_1$ using self-consistency conditions.
Choosing
\begin{equation*}
\mu^2 = D(\varepsilon)^{-1/\varepsilon} (-2\omega)^2 \to e^{-\frac{5}{3}\varepsilon} (-2\omega)^2
\end{equation*}
we have
\begin{equation}
V(\omega,\omega;\varphi) - V(\omega,\omega;0) = \frac{1}{\beta_0}
\sum_{L=1}^\infty \frac{\bar{f}(\varepsilon,L\varepsilon;\varphi)}{L}
\left(\frac{b}{\varepsilon+b}\right)^L
+ \mathcal{O}\left(\frac{1}{\beta_0^2}\right)\,,
\label{Lb0:Vren}
\end{equation}
where $\bar{f}(\varepsilon,u;\varphi)=f(\varepsilon,u;\varphi)-f(\varepsilon,u;0)$.
We expand in $b$, expand $\bar{f}(\varepsilon,u;\varphi)$ in $\varepsilon$ and $u$
and select only $\varepsilon^{-1}$ terms in order to obtain $Z_1$.
All coefficients but $f_{n0}$ cancel:
\begin{equation*}
Z_1(b;\varphi) = 2 \frac{\varphi \cot\varphi - 1}{\beta_0}
\sum_{n=0}^\infty \frac{\hat{f}_n}{n+1} (-b)^{n+1}\,,
\end{equation*}
where
\begin{equation*}
\bar{f}(\varepsilon,0;\varphi) = - 2 \hat{f}(\varepsilon)
(\varphi \cot\varphi - 1)\,,\quad
\hat{f}(\varepsilon) = \sum_{n=0}^\infty \hat{f}_n \varepsilon^n\,.
\end{equation*}
Therefore at the L$\beta_0$ we obtain~\cite{Beneke:1995pq}
\begin{align}
&\Gamma(b;\varphi) = 4 \frac{b}{\beta_0} \gamma_0(b) (\varphi \cot\varphi - 1)
+ \mathcal{O}\left(\frac{1}{\beta_0^2}\right)\,,
\nonumber\\
&\gamma_0(b) = \hat{f}(-b) = \frac{(1+\frac{2}{3}b) \Gamma(2+2b)}{(1+b) \Gamma^3(1+b) \Gamma(1-b)}
\nonumber\\
&{} = 1 + \frac{5}{3} b - \frac{1}{3} b^2
- \left( 2 \zeta_3 - \frac{1}{3} \right) b^3
+ \left( \frac{\pi^4}{30} - \frac{10}{3} \zeta_3 - \frac{1}{3} \right) b^4
+ \cdots
\label{Lb0:BB}
\end{align}

As a free bonus, we can obtain the HQET field anomalous dimension.
The vertex function $V$ at $\varphi=0$ is related to the HQET propagator $S$ by the Ward identity
\begin{equation}
V(\omega,\omega';0) = \frac{S^{-1}(\omega') - S^{-1}(\omega)}{\omega' - \omega}\,,\quad
V(\omega,\omega;0) = \frac{d S^{-1}(\omega)}{d\omega}\,.
\label{Lb0:Ward}
\end{equation}
Therefore the renormalization constant of the HQET quark field $Z_h$ is given by
\begin{equation*}
\log V(\omega,\omega';0) = - \log Z_h + \text{finite}\,.
\end{equation*}
Using
\begin{equation*}
f(\varepsilon,u;0) = - 3
\frac{(1-\frac{2}{3}\varepsilon)^2 \Gamma(2-2\varepsilon) \Gamma(1-u) \Gamma(1+2u)}%
{(1-\varepsilon) \Gamma^2(1-\varepsilon) \Gamma(1+\varepsilon) \Gamma(2+u-\varepsilon)}\,,
\end{equation*}
we obtain in the Landau gauge~\cite{Broadhurst:1994se}
\begin{align}
&\gamma_h(b) = 2 \frac{b}{\beta_0} \gamma_{h0}(b)
+ \mathcal{O}\left(\frac{1}{\beta_0^2}\right)\,,
\nonumber\\
&\gamma_{h0}(b) = f(-b,0;0)
= \frac{\left(1+\frac{2}{3}b\right)^2 \Gamma(2+2b)}{(1+b)^2 \Gamma^3(1+b) \Gamma(1-b)}
\nonumber\\
&{} = 1 + \frac{4}{3} b - \frac{5}{9} b^2
- \left( 2 \zeta_3 - \frac{2}{3} \right) b^3
+ \left( \frac{\pi^4}{30} - \frac{8}{3} \zeta_3 - \frac{7}{9} \right) b^4
+ \cdots
\label{Lb0:BG}
\end{align}
(in an arbitrary covariant gauge, a one-loop gauge-dependent contribution should be added).

Now we consider the potential $V(\vec{q}^{\,})$ at the L$\beta_0$ order.
Choosing $\mu^2=\vec{q}^{\,2}$ we have
\begin{equation*}
V(\vec{q}^{\,}) = - \frac{(4\pi)^{D/2} e^{\gamma \varepsilon}}{\beta_0 D(\varepsilon) (\vec{q}^{\,2})^{1-\varepsilon}}
\varepsilon \sum_{L=1}^\infty \left(D(\varepsilon) \frac{b}{\varepsilon+b}\right)^L
+ \mathcal{O}\left(\frac{1}{\beta_0^2}\right)\,.
\end{equation*}
The sum here can be written as
\begin{equation*}
\sum_{L=1}^\infty g(\varepsilon,L\varepsilon) \left(\frac{b}{\varepsilon+b}\right)^L\,,\quad
g(\varepsilon,u) = D(\varepsilon)^{u/\varepsilon} = \sum_{n,m=0}^\infty g_{nm} \varepsilon^n u^m\,.
\end{equation*}
This sum is equal to
\begin{equation*}
\frac{b}{\varepsilon} \sum_{n=0}^\infty n!\,g_{0n} b^n + \mathcal{O}(\varepsilon^0)
\end{equation*}
($1/\varepsilon^n$ terms with $n>1$ vanish, so that $V(\vec{q}^{\,})$ is automatically finite),
where
\begin{equation}
g(0,u) = e^{\frac{5}{3}u}\,,\quad
g_{0n} = \frac{1}{n!} \left(\frac{5}{3}\right)^n\,.
\label{Lb0:fu}
\end{equation}
Therefore
\begin{equation}
V(\vec{q}^{\,}) = - \frac{(4\pi)^2}{\vec{q}^{\,2}} \frac{b}{\beta_0} V_0(b)
+ \mathcal{O}\left(\frac{1}{\beta_0^2}\right)\,,\quad
V_0(b) = \frac{1}{1 - \frac{5}{3} b}\,.
\label{Lb0:V}
\end{equation}
The conformal anomaly~(\ref{Intro:Delta}) at the L$\beta_0$ order is
\begin{align}
&\Delta = 4 \pi \frac{b^3}{\beta_0} \delta_0(b) + \mathcal{O}\left(\frac{1}{\beta_0^2}\right)\,,
\nonumber\\
&\delta_0(b) = \frac{V_0(b) - \gamma_0(b)}{b^2}
= \frac{28}{9}
+ 2 \left( \zeta_3 + \frac{58}{27} \right) b
- \frac{1}{3} \left( \frac{\pi^4}{10} - 10 \zeta_3 - \frac{652}{27} \right) b^2
+ \cdots
\label{Lb0:Delta}
\end{align}
The first term here reproduces the $T_F n_f$ term in~(\ref{Intro:Delta3}).

\section{Next to leading $\beta_0$ order}
\label{S:NLb0}

To obtain the photon propagator with the NL$\beta_0$ accuracy,
we need the photon self-energy up to $1/\beta_0$:
\begin{equation}
\Pi(k^2)
= \raisebox{-5.5mm}{\includegraphics{pi0.pdf}}
+ 2 \raisebox{-5.5mm}{\includegraphics{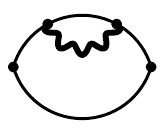}}
+ \raisebox{-5.5mm}{\includegraphics{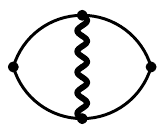}}
= \Pi_0(k^2) + \frac{\Pi_1(k^2)}{\beta_0}
+ \mathcal{O}\left(\frac{1}{\beta_0^2}\right)\,,
\label{NLb0:Pi}
\end{equation}
where the photon propagators in $\Pi_1$ are taken at the L$\beta_0$ order.
The NL$\beta_0$ contribution can be written in the form~\cite{PalanquesMestre:1983zy,Broadhurst:1992si}
\begin{equation}
\Pi_1(k^2) = 3 \varepsilon \sum_{L=2}^\infty \frac{F(\varepsilon,L\varepsilon)}{L} \Pi_0(k^2)^L\,.
\label{NLb0:Pi1}
\end{equation}
Using integration by parts, one can reduce it to
\begin{align}
&F(\varepsilon,u) =
\frac{2 (1-2\varepsilon)^2 (3-2\varepsilon) \Gamma^2(1-2\varepsilon)}%
{9 (1-\varepsilon) (1-u) (2-u) \Gamma^2(1-\varepsilon) \Gamma^2(1+\varepsilon)}
\nonumber\\
&{}\times\biggl[ - u
\frac{2-3\varepsilon-\varepsilon^2 + \varepsilon(2+\varepsilon)u - \varepsilon u^2}{\Gamma^2(1-\varepsilon)}
I(1+u-2\varepsilon)
\nonumber\\
&\hphantom{{}\times\biggl[\biggr.}{} + 2
\frac{2(1+\varepsilon)(3-2\varepsilon) - (4+11\varepsilon-7\varepsilon^2)u + \varepsilon(8-3\varepsilon)u^2 - \varepsilon u^3}%
{(1-u) (2-u) (1-u-\varepsilon) (2-u-\varepsilon)}
\frac{\Gamma(1+u) \Gamma(1-u+\varepsilon)}{\Gamma(1-u-\varepsilon) \Gamma(1+u-2\varepsilon)}
\biggr]
\nonumber\\
&{} = \sum_{n,m=0}^\infty F_{nm} \varepsilon^n u^m\,,
\label{NLb0:Feu}
\end{align}
where the integral
\begin{equation*}
I(n) =
\raisebox{-5.5mm}{\begin{picture}(16,12.5)
\put(8,6.25){\makebox(0,0){\includegraphics{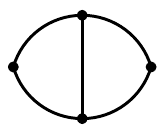}}}
\put(9.5,6.25){\makebox(0,0){$n$}}
\end{picture}}
= \frac{1}{\pi^d} \int \frac{d^d k_1\,d^d k_2}%
{k_1^2 k_2^2 (k_1+p)^2 (k_2+p)^2 \left[(k_1-k_2)^2\right]^n}
\end{equation*}
(euclidean, $p^2=1$) can be expressed via a ${}_3F_2$ function of unit argument~\cite{Kotikov:1995cw,Broadhurst:1996ur}
(see the review~\cite{Grozin:2012xi} for more references).
The ${}_3F_2$ function can be expanded up to any desired order using known algorithms,
the coefficients are expressed via multiple $\zeta$ values;
therefore, the coefficients $F_{nm}$ can be calculated to any desired order.

The function $F(\varepsilon,u)$ simplifies in some cases.
In particular~\cite{PalanquesMestre:1983zy},
\begin{equation}
F(\varepsilon,0) =
\frac{(1+\varepsilon) (1-2\varepsilon)^2 (1-\frac{2}{3}\varepsilon)^2 \Gamma(1-2\varepsilon)}%
{(1-\varepsilon)^2 (1-\frac{1}{2}\varepsilon) \Gamma(1+\varepsilon) \Gamma^3(1-\varepsilon)}\,,
\label{NLb0:Fe0}
\end{equation}
so that $F_{n0}$ contain no multiple $\zeta$ values, only $\zeta_n$.
Also~\cite{Broadhurst:1992si}
\begin{equation}
F(0,u) = \frac{2}{3}
\frac{\psi'\left(2-\frac{u}{2}\right) - \psi'\left(1+\frac{u}{2}\right)
- \psi'\left(\frac{3-u}{2}\right) + \psi'\left(\frac{1+u}{2}\right)}%
{(1-u) (2-u)}
\label{NLb0:F0u}
\end{equation}
so that $F_{0m}$ contains only $\zeta_{2n+1}$~\cite{Broadhurst:1992si}:
\begin{equation}
F_{0m} = - \frac{32}{3} \sum_{s=1}^{[(m+1)/2]} s \left(1 - 2^{-2s}\right) \left(1 - 2^{2s-m-2}\right) \zeta_{2s+1}
+ \frac{4}{3} (m+1) \left(m + (m+6) 2^{-m-3}\right)\,.
\label{NLb0:F0m}
\end{equation}
The two-loop case is, of course, trivial:
\begin{equation*}
F(\varepsilon,2\varepsilon) = \frac{2}{9\varepsilon^2} \frac{3-2\varepsilon}{1-\varepsilon}
\biggl[ 2 \frac{(1-2\varepsilon)^2 (2-2\varepsilon+\varepsilon^2)}{(1-3\varepsilon) (2-3\varepsilon)}
\frac{\Gamma(1+2\varepsilon) \Gamma^2(1-2\varepsilon)}{\Gamma^2(1+\varepsilon) \Gamma(1-\varepsilon) \Gamma(1-3\varepsilon)}
- 2 + \varepsilon - 2 \varepsilon^2 \biggr]\,.
\end{equation*}

Let's write the charge renormalization constant $Z_\alpha$ with the NL$\beta_0$ accuracy as
\begin{align}
&Z_\alpha(b) = \frac{1}{1+b/\varepsilon}
\left[1 + \frac{Z_{\alpha1}(b)}{\beta_0} + \mathcal{O}\left(\frac{1}{\beta_0^2}\right) \right]\,,
\nonumber\\
&Z_{\alpha1}(b) = \frac{Z_{\alpha11}(b)}{\varepsilon} + \frac{Z_{\alpha12}(b)}{\varepsilon^2} + \cdots\,,\quad
Z_{\alpha1n} = \mathcal{O}(b^{n+1})\,.
\label{NLb0:Za}
\end{align}
In the abelian theory, $\log(1-\Pi)$ expressed~(\ref{Lb0:MS}) via renormalized $b$
should be equal to $\log Z_\alpha + \text{finite}$.
Equating the coefficients of $\varepsilon^{-1}$ in the $1/\beta_0$ terms in this relation,
we see that $Z_{\alpha11}$~(\ref{NLb0:Za}) is given by the coefficient of $\varepsilon^{-1}$ in
\begin{equation*}
- \left(1 + \frac{b}{\varepsilon}\right) \Pi_1\,.
\end{equation*}
It is convenient to choose
\begin{equation*}
\mu^2 = D(\varepsilon)^{-1/\varepsilon} (-k^2) \to e^{-\frac{5}{3}\varepsilon} (-k^2)\,,
\end{equation*}
then
\begin{equation*}
\Pi_1 = 3 \varepsilon \sum_{L=2}^\infty \frac{F(\varepsilon,L\varepsilon)}{L}
\left(\frac{b}{\varepsilon+b}\right)^L\,.
\end{equation*}
We expand in $b$ and expand $F(\varepsilon,u)$ in $\varepsilon$ and $u$;
selecting $\varepsilon^{-1}$ terms, we find that all coefficients but $F_{n0}$ cancel:
\begin{equation}
Z_{\alpha11} = - 3 \sum_{n=0}^\infty \frac{F_{n0} (-b)^{n+2}}{(n+1) (n+2)}\,.
\label{NLb0:Z1}
\end{equation}
The $\beta$ function with NL$\beta_0$ accuracy is
\begin{equation}
\beta(b) = b + \frac{\beta_1(b)}{\beta_0} + \mathcal{O}\left(\frac{1}{\beta_0^2}\right)\,,
\label{NLb0:beta}
\end{equation}
where~\cite{PalanquesMestre:1983zy,Broadhurst:1992si}
\begin{align}
&\beta_1(b) = - \frac{d Z_{\alpha11}(b)}{d\log b}
= 3 \sum_{n=0}^\infty \frac{F_{n0} (-b)^{n+2}}{n+1}
\nonumber\\
&{} = 3 b^2 + \frac{11}{4} b^3 - \frac{77}{36} b^4
- \frac{1}{2} \left( 3 \zeta_3 + \frac{107}{48} \right) b^5
+ \frac{1}{5} \left( \frac{\pi^4}{10} - 11 \zeta_3 + \frac{251}{48} \right) b^6
+ \cdots
\label{nf:beta}
\end{align}
(the coefficients $F_{n0}$ follow from $F(\varepsilon,0)$~(\ref{NLb0:Fe0})).
The corresponding terms in the 5-loop QED $\beta$ function~\cite{Baikov:2012zm} are reproduced.
We shall need the full $Z_{\alpha1}$, not just $Z_{\alpha11}$;
integrating the RG equation with the $1/\beta_0$ accuracy we obtain
\begin{equation*}
Z_{\alpha1}(b) = - \varepsilon \int_0^b \frac{\beta_1(b)\,d b}{b (\varepsilon+b)^2}
= - \frac{3}{2} \frac{b^2}{\varepsilon}
+ \frac{1}{2} \left(4 + F_{10} \varepsilon \right) \frac{b^3}{\varepsilon^2}
- \frac{1}{4} \left(9 + 3 F_{10} \varepsilon + F_{20} \varepsilon^2 \right) \frac{b^4}{\varepsilon^3}
+ \cdots
\end{equation*}

\begin{figure}[b]
\begin{center}
\includegraphics{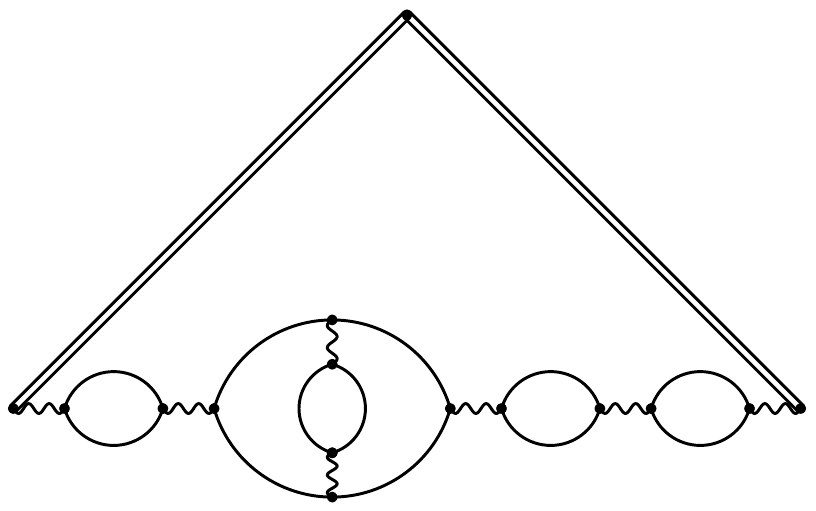}
\end{center}
\caption{NL$\beta_0$ order diagrams contain one $\Pi_1$ insertion (with any number of $\Pi_0$ insertions inside)
and any number of $\Pi_0$ insertions to the left and to the right of it.}
\label{F:NLb0}
\end{figure}

At the NL$\beta_0$ order we should expand
the photon propagator $(1-\Pi_0-\Pi_1/\beta_0)^{-1}$ up to $1/\beta_0$ (Fig.~\ref{F:NLb0}).
The vertex function~(\ref{Lb0:Vren}) becomes
\begin{align}
&V(\omega,\omega;\varphi) - V(\omega,\omega;0)
= \frac{1}{\beta_0} \sum_{L=1}^\infty \frac{\bar{f}(\varepsilon,L\varepsilon;\varphi)}{L}
\left(\frac{b}{\varepsilon+b}\right)^L
\nonumber\\
&{}\times\left[1 + L \frac{Z_{\alpha1}}{\beta_0}
+ \frac{3\varepsilon}{\beta_0} \sum_{L'=2}^{L-1} \frac{L-L'}{L'} F(\varepsilon,L'\varepsilon) \right]
+ \mathcal{O}\left(\frac{1}{\beta_0^3}\right)\,,
\label{NLb0:Vw}
\end{align}
where $L'$ is the number of loops in the $\Pi_1$ insertion,
and the $1/\beta_0$ correction $Z_{\alpha1}$ to the charge renormalization~(\ref{NLb0:Za})
is taken into account.
We expand in $b$ and substitute the expansions~(\ref{NLb0:Feu}) and~(\ref{Lb0:feuexp});
in $Z_1$, the coefficient of $\varepsilon^{-1}$, all $\bar{f}_{nm}$ except $\bar{f}_{n0}$ cancel.
At the NL$\beta_0$ order the cusp anomalous dimension is determined
by the same $\hat{f}_n$ coefficients as at the L$\beta_0$ order:
\begin{equation}
\Gamma(b;\varphi) = 4 \left[ \frac{b}{\beta_0} \gamma_0(b) - \frac{b^3}{\beta_0^2} \gamma_1(b) \right]
\left( \varphi \cot\varphi - 1 \right) + \mathcal{O}\left(\frac{1}{\beta_0^3}\right)\,,
\label{NLb0:struct}
\end{equation}
where
\begin{align*}
&\gamma_1(b) = - \frac{3}{2} \bigl[F_{10} + 2 F_{01} - 2 \hat{f}_1 \bigr]
+ \bigl[2 F_{20} + 3 (F_{11} + F_{02}) + 3 F_{01} \hat{f}_1 - 6 \hat{f}_2 \bigr] b\\
&{}
- \biggl[ \frac{3}{4} (3 F_{30} + 4 (F_{21} + F_{12} + F_{03}))
+ (F_{20} + 3 (F_{11} + F_{02})) \hat{f}_1
- \frac{3}{2} \bigl(F_{10} - 2 F_{01}\bigr) \hat{f}_2
- 9 \hat{f}_3 \biggr] b^2
+ \cdots
\end{align*}
Substituting $F_{nm}$ we obtain
\begin{align}
&\gamma_1(b) = 12 \zeta_3 - \frac{55}{4}
+ \biggl( - \frac{\pi^4}{5} + 40 \zeta_3 - \frac{299}{18} \biggr) b
\nonumber\\
&{}+ \biggl( 24 \zeta_5 - \frac{2}{3} \pi^4 + \frac{233}{6} \zeta_3 + \frac{15211}{864} \biggr) b^2
\nonumber\\
&{}+ \biggl( - 48 \zeta_3^2 - \frac{2}{63} \pi^6 + 80 \zeta_5 - \frac{167}{225} \pi^4 + \frac{1168}{15} \zeta_3 - \frac{971}{240} \biggr) b^3
\nonumber\\
&{}+ \biggl( 36 \zeta_7 + \frac{8}{5} \pi^4 \zeta_3 - 160 \zeta_3^2 - \frac{20}{189} \pi^6 + \frac{377}{3} \zeta_5 - \frac{23}{15} \pi^4
+ \frac{929}{12} \zeta_3 - \frac{8017}{1728} \biggr) b^4
\nonumber\\
&{}+ \biggl( - 240 \zeta_3 \zeta_5 - \frac{4}{225} \pi^8 + 120 \zeta_7 + \frac{16}{3} \pi^4 \zeta_3 - \frac{2776}{21} \zeta_3^2 - \frac{914}{3969} \pi^6
\nonumber\\
&\hphantom{{}+\biggl(\biggr.}{}
+ \frac{6826}{21} \zeta_5 - \frac{1793}{1350} \pi^4 - \frac{31693}{315} \zeta_3 + \frac{79433}{4320} \biggr) b^5
+ \cdots
\label{NLb0:Gamma}
\end{align}
This expansion can be extended to any number of loops.
The first term in~(\ref{NLb0:Gamma}) agrees with the $C_F^2 T_F n_f$ term
in the three-loop result~\cite{Grozin:2014axa,Grozin:2014hna,Grozin:2015kna}.
The next term coincides with the $C_F^2 (T_F n_f)^2 \alpha_s^4$ term in $\Gamma$
recently calculated in~\cite{Vogt}.
Note that the last (8-loop) term here contains $F_{nm}$ with $n+m=6$, $n>0$, $m>0$, which contain $\zeta_{5,3}$;
but they enter as the combination $F_{51}+F_{42}+F_{33}+F_{24}+F_{15}$ in which this $\zeta_{5,3}$ cancels.

Similarly, the field anomalous dimension in Landau gauge at the NL$\beta_0$ order is
\begin{align}
&\gamma_h(b) = - 6 \left[ \frac{b}{\beta_0} \gamma_{h0}(b) - \frac{b^3}{\beta_0^2} \gamma_{h1}(b) \right]
+ \mathcal{O}\left(\frac{1}{\beta_0^3}\right)\,,
\nonumber\\
&\gamma_{h1}(b) = 3 \biggl( 4 \zeta_3 - \frac{17}{4} \biggr)
+ \biggl( - \frac{\pi^4}{5} + 36 \zeta_3 - \frac{103}{9} \biggr) b
\nonumber\\
&{} + \biggl( 24 \zeta_5 - \frac{3}{5} \pi^4 + \frac{59}{2} \zeta_3 + \frac{14579}{864} \biggr) b^2
\nonumber\\
&{} + \biggl( - 48 \zeta_3^3 - \frac{2}{63} \pi^6 + 72 \zeta_5 - \frac{44}{75} \pi^4 + \frac{3229}{45} \zeta_3 - \frac{5191}{540} \biggr) b^3
\nonumber\\
&{} + \biggl( 36 \zeta_7 + \frac{8}{5} \pi^4 \zeta_3 - 144 \zeta_3^2 - \frac{2}{21} \pi^6 + 107 \zeta_5 - \frac{946}{675} \pi^4
+ \frac{9601}{180} \zeta_3 + \frac{22859}{8640} \biggr) b^4
\nonumber\\
&{} + \biggl( - 240 \zeta_3 \zeta_5 - \frac{4}{225} \pi^8 + 108 \zeta_7 + \frac{24}{5} \pi^4 \zeta_3 - \frac{664}{7} \zeta_3^2  - \frac{272}{1323} \pi^6
\nonumber\\
&\hphantom{{}+\biggl(\biggr.}{}
+ \frac{18574}{63} \zeta_5 - \frac{119}{135} \pi^4 - \frac{6263}{63} \zeta_3 + \frac{16103}{1296}
\biggr) b^5 + \cdots
\label{NLb0:gammah}
\end{align}
The first term here coincides with the $C_F^2 T_F n_f$ term in the three-loop result
obtained by a direct calculation~\cite{Melnikov:2000zc,Chetyrkin:2003vi}.
The last term contains the same combination of $F_{nm}$ with $n+m=6$,
so that $\zeta_{5,3}$ cancels.

The static potential at the NL$\beta_0$ level is
\begin{align}
V(\vec{q}^{\,}) &{}= - \frac{(4\pi)^2}{\beta_0 \vec{q}^{\,2}} \varepsilon
\sum_{L=1}^\infty g(\varepsilon,L\varepsilon) \left(\frac{b}{\varepsilon+b}\right)^L
\left[1 + L \frac{Z_{\alpha1}}{\beta_0} + \frac{3 \varepsilon}{\beta_0}
\sum_{L'=2}^{L-1} \frac{L-L'}{L'} F(\varepsilon,L'\varepsilon) \right]
+ \mathcal{O}\left(\frac{1}{\beta_0^3}\right)
\nonumber\\
&{}= - \frac{(4\pi)^2}{\vec{q}^{\,2}}
\left[ \frac{b}{\beta_0} V_0(b) - \frac{b^3}{\beta_0^2} V_1(b) \right]
+ \mathcal{O}\left(\frac{1}{\beta_0^3}\right)
\label{NLb0:V}
\end{align}
where
\begin{align*}
&V_1(b) = - \frac{3}{2} \left[ F_{10} + 2 F_{01} + 2 g_{01} \right]
+ \frac{1}{2} \left[ F_{20} - 6 F_{02} - 6 \left( F_{10} + 3 F_{01} \right) g_{01} - 30 g_{02} \right] b\\
&{} - \frac{1}{4} \bigl[ F_{30} + 24 F_{03} - 4 \left( F_{20} + 12 F_{02} \right) g_{01}
+ 36 \left( F_{10} + 4 F_{01} \right) g_{02} + 312 g_{03} \bigr] b^2 + \cdots
\end{align*}
contains only the same coefficients $g_{0n}$~(\ref{Lb0:fu}) as the L$\beta_0$ result,
and only $F_{n0}$ and $F_{0m}$ are involved (see~(\ref{NLb0:Fe0}--\ref{NLb0:F0m})).
We obtain
\begin{align}
&V_1(b) = 12 \zeta_3 - \frac{55}{4}
+ \left( 78 \zeta_3 - \frac{7001}{72} \right) b
+ \left( 60 \zeta_5 + \frac{723}{2} \zeta_3 - \frac{147851}{288} \right) b^2
\nonumber\\
&{} + \left( 770 \zeta_5 + \frac{\pi^4}{200} + \frac{276901}{180} \zeta_3 - \frac{70418923}{25920} \right) b^3
\nonumber\\
&{} + \left( 1134 \zeta_7 + \frac{32297}{5} \zeta_5 + \frac{41}{1800} \pi^4 + \frac{402479}{60} \zeta_3
- \frac{1249510621}{77760} \right) b^4
\nonumber\\
&{} + \biggl( 21735 \zeta_7 + \frac{\zeta_3^2}{7} + \frac{\pi^6}{1323} + \frac{5911849}{126} \zeta_5 + \frac{41}{720} \pi^4
+ \frac{48558187}{1512} \zeta_3 - \frac{10255708489}{93312} \biggr) b^5
+ \cdots
\label{NLb0:V1}
\end{align}
Thus we have reproduced the $C_F (T_F n_f)^2 \alpha_s^3$ and $C_F^2 T_F n_f \alpha_s^3$ terms in the two-loop potential~\cite{Schroder:1998vy},
as well as the $C_F (T_F n_f)^3 \alpha_s^4$ and $C_F^2 (T_F n_f)^2 \alpha_s^4$ terms in the three-loop one~\cite{Smirnov:2008pn}.
This expansion can be extended to any order;
it contains only $\zeta_n$ because only $F_{n0}$ and $F_{0m}$ are present.
Note the pattern of the highest weights in~(\ref{NLb0:V1}): 3, 3, 5, 5, 7, 7,
whereas one would expect 3, 4, 5, 6, 7, 8, as in~(\ref{NLb0:Gamma}), (\ref{NLb0:gammah}).
The conformal anomaly~(\ref{Intro:Delta}) at the NL$\beta_0$ order is
\begin{align}
&\Delta = 4 \pi \left[ \frac{b^3}{\beta_0} \delta_0(b) - \frac{b^4}{\beta_0^2} \delta_1(b) \right]
+ \mathcal{O}\biggl(\frac{1}{\beta_0^3}\biggr)\,,
\nonumber\\
&\delta_1(b) = \frac{\pi^4}{5} + 38 \zeta_3 - \frac{645}{8}
+ \biggl( 36 \zeta_5 + \frac{2}{3} \pi^4 + \frac{968}{3} \zeta_3 - \frac{114691}{216} \biggr) b
\nonumber\\
&{} + \biggl( 48 \zeta_3^2+ \frac{2}{63} \pi^6 + 690 \zeta_5 + \frac{269}{360} \pi^4
+ \frac{52577}{36} \zeta_3 - \frac{14062811}{5184} \biggr) b^2
\nonumber\\
&{} + \biggl( 1098 \zeta_7 - \frac{8}{5} \pi^4 \zeta_3 + 160 \zeta_3^2 + \frac{20}{189} \pi^6 + \frac{95006}{15} \zeta_5 + \frac{2801}{1800} \pi^4
+ \frac{198917}{30} \zeta_3 - \frac{39035933}{2430} \biggr) b^3
\nonumber\\
&{} + \biggl( 240 \zeta_3 \zeta_5 + \frac{4}{225} \pi^8 + 21615 \zeta_7 - \frac{16}{3} \pi^4 \zeta_3
+ \frac{397}{3} \zeta_3^2 + \frac{131}{567} \pi^6
\nonumber\\
&\hphantom{{}+\biggl(\biggr.}{}
+ \frac{838699}{18} \zeta_5 + \frac{14959}{10800} \pi^4 + \frac{34793081}{1080} \zeta_3 - \frac{51287121209}{466560} \biggr) b^4
+ \cdots
\label{NLb0:Delta}
\end{align}
The $b^3/\beta_0^2$ term has canceled, so that the coefficient of $C_F$ in the bracket in~(\ref{Intro:Delta3}) is 0.

\section{Conclusion}
\label{S:Conc}

The terms with the highest powers of $n_f$ at each order of perturbation theory
($C_F (T_F n_f)^{L-1} \alpha_s^L$ in $\Gamma$, $\gamma_h$; $C_F (T_F n_f)^L \alpha_s^{L+1}$ in $V(\vec{q}^{\,})$)
are known, and given by explicit formulas~(\ref{Lb0:BB}), (\ref{Lb0:BG}), (\ref{Lb0:V}).
The terms with the next to highest power of $n_f$ can have abelian ($C_F^2$) or non-abelian ($C_F C_A$) color structure.
The abelian terms
($C_F^2 (T_F n_f)^{L-2} \alpha_s^L$ ($L\ge3$) in $\Gamma$, $\gamma_h$;
$C_F^2 (T_F n_f)^{L-1} \alpha_s^{L+1}$ ($L\ge2$) in $V(\vec{q}^{\,})$)
are also known to all orders in $\alpha_s$,
but only as algorithms which allow one to obtain (in principle) any number of terms,
see~(\ref{NLb0:Gamma}), (\ref{NLb0:gammah}), (\ref{NLb0:V1}).
The simple method used here is not applicable to non-abelian terms.

I am grateful to J.\,M.~Henn, G.\,P.~Korchemsky, P.~Marquard for collaboration~\cite{Grozin:2014axa,Grozin:2014hna,Grozin:2015kna};
to D.\,J.~Broadhurst for explaining the methods of~\cite{Broadhurst:1992si} and useful discussions;
to A.~Vogt for comparing the result~\cite{Vogt} with~(\ref{NLb0:Gamma}) during the conference.
Many thanks to the organizers of Loops and Legs 2016.
I am grateful to MITP and M.~Neubert for hospitality in Mainz and financial support.
Partial support from the Russian Ministry of Education and Science is acknowledged.

\end{document}